\documentclass[aps,pre,twocolumn,superscriptaddress]{revtex4}
\usepackage{amsmath,amssymb}\usepackage{graphicx}\usepackage{subfigure}\usepackage{color}\usepackage{psfig}
\usepackage{times}\usepackage{tabularx}\usepackage{color}
\begin{document} \title{\bf {Rigorous error bounds for Ewald summation of electrostatics at planar intefaces}}
\author{Cong Pan}
%\affiliation{State Key Laboratory of Supramolecular Structure and Materials,Institute of Theoretical Chemistry, Jilin University, Changchun, 130012, P. R. China}
\affiliation{State Key Laboratory of Supramolecular Structure and Materials, Jilin University, Changchun, 130012, P. R. China}
\affiliation{Institute of Theoretical Chemistry, Jilin University, Changchun, 130012, P. R. China}
\author{Zhonghan Hu} \email{zhonghanhu@jlu.edu.cn}
\affiliation{State Key Laboratory of Supramolecular Structure and Materials, Jilin University, Changchun, 130012, P. R. China}
\affiliation{Institute of Theoretical Chemistry, Jilin University, Changchun, 130012, P. R. China}
\affiliation{Kavli Institute for Theoretical Physics China, CAS, Beijing 100190, P. R. China}
\date{\today}\normalsize

\begin{abstract}
We present a rigorous Ewald summation formula to evaluate the electrostatic interactions in two-dimensionally periodic planar interfaces of three-dimensional
systems. By rewriting the Fourier part of the summation formula of the original Ewald2D expression with an explicit order $N^2$ complexity to a
closed form Fourier integral,
 we find that both the previously developed electrostatic layer correction term and the boundary correction term naturally arise from the expression of
a rigorous trapezoidal summation of the Fourier integral part. We derive the exact corrections to the trapezoidal summation in a form of contour integrals offering
precise error bounds with given parameter sets of mesh size and system length. Numerical calculations of Madelung constants in model ionic crystals of slab geometry
have been performed to support our analytical results.
\end{abstract}

\maketitle

%%%%%%%%%%%%%%%%%%%%%%%%%%%%%%%%%%%%%%%%%%%%%%%%%%%%%%%%%%%%%%%%%%%%%%%%%%%%%%%%%%%%%%%%%%%%%%%%%%%%%%%%%%%%%%%%%%%%%%%%%%%%%%%
\section{Introduction}
\begin{figure}
\centerline{
\psfig{file=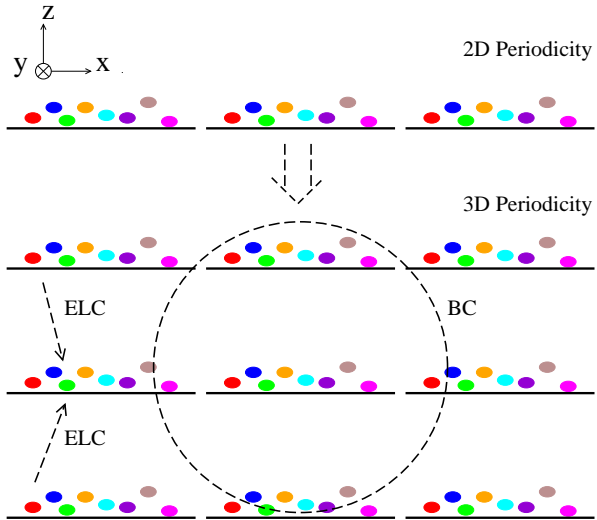,width=3.2in,angle=270}}
\caption{Illustration of the systems of interest with 2D (up) and 3D (down) periodicity. $x$, $y$, and $z$ directions are pointing to the right, into the paper, and
up respectively. When transferring the evaluation of the electrostatic interaction in a system with 2D periodicity to that of 3D periodicity, the boundary correction term
and the electrostatic layer correction term are added.}
\label{fig:period2D}
\end{figure}

Structural and dynamical properties of liquid-vapor, liquid-liquid, and liquid-solid interfaces are of great interest to chemists, physicists, and material
scientists\cite{giovambattistaREV,benjaminREV,davidREV,takeshiREV}. As atoms in the molecules generally have partial charges, determination of the interfacial properties requires an accurate treatment of the long-ranged
Coulomb interactions in a slab geometry of two-dimensional (2D) periodicity. However, it remains a great challenge for theoretical physical chemists to efficiently
and accurately treat the long-ranged Coulomb interaction when performing molecular dynamics or Monte Carlo simulation studies for three-dimensional (3D) charged
systems with 2D (e.g. $x$ and $y$ directions) periodic boundary condition (PBC) and the other dimension (e.g. $z$ direction) nonperiodic (see
Fig.~\ref{fig:period2D}).
More than 30 years ago, an Ewald summation method for 3D charged systems with 2D periodicity was first derived by Parry\cite{Parry1975}, by Heyes, Barber and
Clarker\cite{Heyes_Clarke1977} and by de Leeuv and Perram\cite{DeLeeuw_Perram1979} (referred as Ewald2D method). For a system of $N$ charged particles, the
computational cost of Ewald2D method carried out over distinct pairs of charges scales as ${\cal O}(N^2)$ and becomes very expensive when $N$ is very large. Many
researchers have thus proposed alternative Ewald type methods to introduce significant reduction in the computational time
\cite{Hautman_Klein1992,Widmann_Adolf1997,Spohr1997,Yeh_Berkowitz1999,Grzybowski_Brodka2000,Kawata_Mikami2001,Arnold_Holm2002,Arnold_Holm2002b,Tuckerman_Martyna2002,Lindbo_Tornberg2012}.
Hautmann and Klein developed a method based on an expansion of Ewald2D expression to a Taylor series which can be applied to thin layer configurations where the extension in the dimension with no periodicity (length in $z$ direction) is much smaller than the length of the unit cell in the other two
dimensions ($x$ and $y$ directions)\cite{Hautman_Klein1992,Widmann_Adolf1997}.
For a system of charges with an arbitrary length in the nonperiodic dimension, the first successful method to improve the efficiency of Ewald type computation with
sufficient accuracy was proposed by Yeh and Berkowitz\cite{Yeh_Berkowitz1999}. Their method relates the electrostatic potential of a 3D system of 2D PBC to that of a
corresponding 3D system of 3D PBC but with a correction term accounting for the difference between the two boundary conditions (refered as Ewald3DC method)
\cite{Spohr1997,Yeh_Berkowitz1999}. Sufficient accuracy of Ewald3DC method can be always obtained by introducing large empty space in the artificial 3D periodic
system. The Ewald3DC method has been widely applied to molecular dynamics or Monte Carlo simulation studies of various interfacial systems in slab geometry (e.g.
ref.\cite{Hu_Weeks2010a}).
With the same intuition of relating electrostatics in 2D periodicity to that in 3D periodicity (see Fig.~\ref{fig:period2D}), Holm and coworkers have further added
an electrostatic layer correction term to remove the effect of artificial electrostatic layers in the corresponding 3D full periodic system (referred as EwaldELC
method)\cite{Arnold_Holm2002b}. 
Aside from extending the problem to full 3D periodicity and then correcting the effect of boundary and electrostatic layers, several researchers have sought to
rewrite the Ewald2D expression in an alternative way in order to obtain better computational scaling
\cite{Kawata_Mikami2001,Tuckerman_Martyna2002,Lindbo_Tornberg2012}.
Notably, recent development by Lindbo and Tornberg have successfully developed fast and accurate Ewald2D techniques which involves a Particle Mesh type
generalization\cite{Lindbo_Tornberg2012}. These types of Ewald2D summation methods by Kawata and Mikami\cite{Kawata_Mikami2001} or Lindbo and
Tornberg\cite{Lindbo_Tornberg2012} start from an equivalent Fourier integral expression of the Ewald2D formula and apply an interpolation treatment for a term
purely depending on $z$ and a fast griding treatment for the other part of the Fourier integral. The above development has focused on the exact treatment of the
long-ranged Coulomb interaction which obviously leads to exact dynamical and structural properties of interfaces. 
On the other hand, Weeks and coworkers have developed a mean field treatment called local molecular field theory
\cite{Weeks2002,Chen_Weeks2004,Rodgers_Weeks2008,Rodgers_Weeks2011} which directly treats the short-ranged part of the Coulomb interaction and maps the effect
of the long-ranged part of the Coulomb interaction to a mean-field single particle external potential. When combined with a configuration-based linear response
theory, simulations based on local molecular field theory are able to yield accurate structural and thermodynamic properties with a significant
reduction in computation cost\cite{Hu_Weeks2010}. However, a controllable way to achieve accurate dynamics has yet to be developed for simulation studies of
interfaces in the framework of local molecular field theory.

To efficiently reduce the computational cost and accurately determine the instantaneous potential and force, the Fourier integral expression of the Ewald2D
formula derived as early as in the work by de Leeuv and Perram\cite{DeLeeuw_Perram1979} has been often used as a starting point in the previous developments
\cite{Kawata_Mikami2001,Tuckerman_Martyna2002,Alireza_Goedecker2007,Lindbo_Tornberg2012}. In the present work, we will start from the same Fourier
integral formula and mathematically analyze how the error can be controlled when approximating the integral formula. 
To the best of our knowledge, the point of novelty in the present work is threefold. First, we will suggest a new formulation for the $z$ dependent term which might
be superior to the previous interpolation methods. Second, we will directly derive the boundary correction and electrostatic layer correction terms from the Fourier
integral expression. Our analytical derivation will make a natural connection to the previous Ewald3DC and EwaldELC method. Third, we will show that the error bound due
to the trapezoidal approximation to the Fourier integral expression in Ewald2D is a Gaussian decay function of an appropriate combination of setup parameters. 
The rest of this paper is organized as follows. 
In section~\ref{sec:bg}, we provide brief background on Ewald2D summation method. 
In sections~\ref{sec:alternative} to~\ref{sec:terms}, we formulate an alternative expression of the $z$ independent term in Ewald2D summation, derive rigorous error
bounds when applying trapezoidal approximation to Fourier integrals, and naturally develop the boundary correction term and the electrostatic layer correction term
respectively. In section~\ref{sec:conclusion}, we draw conclusions from our present work.

%%%%%%%%%%%%%%%%%%%%%%%%%%%%%%%%%%%%%%%%%%%%%%%%%%%%%%%%%%%%%%%%%%%%%%%%%%%%%%%%%%%%%%%%%%%%%%%%%%%%%%%%%%%%%%%%%%%%%%%%%%%%%%%%%%%%%%%%%%%%%%%%%%%%%%%%%%%%%%%%%%%%
\section{General background for Ewald2D method} \label{sec:bg}
Let's consider a unit cell of $N$ charged particles located at positions ${\mathbf r}_1,\cdots,{\mathbf r}_N$ respectively. The lengths of the unit cell in $x$ and
$y$ direction are $L_x$ and $L_y$ respectively. The $j$-th particle has its infinite periodic images in 2D located at positions ${\mathbf r}_j +{\mathbf n}$ where
the vector ${\mathbf n}$ stands for $(n_xL_x, n_yL_y, 0)$ and both $n_x$ and $n_y$ are integers (see Fig.~\ref{fig:period2D}). The total electrostatic potential
energy per unit cell is obtained by adding up all Coulomb interactions:
\begin{equation} \label{eq:uorigin}
{\mathcal U} = \frac{1}{2}\sideset{}{'}\sum_{\mathbf n}\left(\sum_{i=1}^N \sum_{j=1}^N \frac{ \displaystyle q_iq_j }{ \displaystyle \left|
 {\mathbf n} + {\mathbf r}_{ij} \right| } \right), \end{equation}
where ${\mathbf r}_{ij}$ stands for the relative vector between the $i$-th and the $j$-th particle. The sum over the vector ${\mathbf n}$ is the sum over infinite images of
$N$ points in both $x$ and $y$ directions. The prime indicates that the $i=j$ term is omitted in case of $n_x=n_y=0$. For simplicity of notation, we have omitted
the prefactor of $1/(4\pi\epsilon_0)$.

It is known that the series ${\mathcal U}$ of eq.~\eqref{eq:uorigin} is slowly convergent subject to the condition of electroneutrality: $\sum_j q_j=0$.
Because of the slow decay of the Coulomb interaction $1/r$ in the series, the straightforward term-by-term summation in real space is impractical for an accurate
determination of ${\mathcal U}$. The Ewald sum introduces a screening factor $\alpha$ and separates the Coulomb interaction in eq.~\eqref{eq:uorigin} into a
combination of short- and long-ranged components\cite{Ewald1921}:
\begin{equation}\label{eq:temp1invr} \frac{1}{r} = \frac{ {\rm erfc}(\alpha r)}{r} + \frac{ {\rm erf}(\alpha r)}{r}, \end{equation}
where the long-ranged component is proportional to the electrostatic potential arising from a normalized Gaussian charge distribution with width $1/\alpha$,
\begin{equation}\label{eq:errorfunc}
  \frac{ {\rm erf}(\alpha r)}{r} \equiv \frac{\alpha^3}{\pi^{3/2}}\int \,dr^\prime e^{-\alpha^2 {r^\prime}^2}\frac{\displaystyle 1}
{\displaystyle  \left| {\mathbf r} - {\mathbf r}^\prime \right|}. \end{equation}
The sum over the vector ${\mathbf n}$ for the short-ranged component decays very fast and is carried out in real space as is done for the usual Lennard-Jones
potential. One can convert the sum over the vector ${\mathbf n}$ for the long-ranged component to a sum over the reciprocal (Fourier) space
vector ${\mathbf h} = 2\pi(h_x/L_x,h_y/L_y,0)$ via Fourier transform. It has been shown that the series ${\mathcal U}$ of eq.~\eqref{eq:uorigin} subject to the
electroneutrality condition is a combination of real space sum and reciprocal space sum
\cite{Parry1975,Heyes_Clarke1977,DeLeeuw_Perram1979,Tuckerman_Martyna2002,Lindbo_Tornberg2012}:
%\begin{equation*} {\mathcal U} = {\mathcal U}_R + \textcolor{black}{{\mathcal U}_F^h} + \textcolor{black}{{\mathcal U}_F^0} , \end{equation*}
\begin{equation}\label{eq:totalU} {\mathcal U} = {\mathcal U}_R + {\mathcal U}_F^h + {\mathcal U}_F^0 , \end{equation}
where
\begin{equation} \label{eq:uewald2dR}
{\mathcal U}_R =\frac{1}{2}\sum_{i,j}q_iq_j\sideset{}{'}\sum_{\mathbf n} \frac{{\rm erfc}(\alpha|{\mathbf r}_{ij}+{\mathbf n}|)}
                {|{\mathbf r}_{ij}+{\mathbf n}|} - \frac{\alpha}{\sqrt\pi} \sum_i q_i^2,  \end{equation}
%\begin{eqnarray} \label{eq:uewald2dFh}
%{\mathcal U}_F^h=  \frac{\pi}{2L_xL_y}\sum_{i,j}q_iq_j\sum_{{\mathbf h}\neq0} \frac{e^{i{\mathbf h}\cdot{\mathbf r}_{ij}}}{h}
%             \left[ e^{hz_{ij}} {\rm erfc}(\frac{h}{2\alpha}+\alpha z_{ij})  \right. 
%            + \left.e^{-hz_{ij}} {\rm erfc}(\frac{h}{2\alpha}-\alpha z_{ij})  \right],   \end{eqnarray}
\begin{multline} \label{eq:uewald2dFh}
{\mathcal U}_F^h=  \frac{\pi}{2L_xL_y}\sum_{i,j}q_iq_j\sum_{{\mathbf h}\neq0} \frac{e^{i{\mathbf h}\cdot{\mathbf r}_{ij}}}{h}
             \left[ e^{hz_{ij}} {\rm erfc}(\frac{h}{2\alpha}+\alpha z_{ij})  \right. \\
            + \left.e^{-hz_{ij}} {\rm erfc}(\frac{h}{2\alpha}-\alpha z_{ij})  \right],   \end{multline}
and
\begin{equation} \label{eq:uewald2dF0}
{\mathcal U}_F^0=\frac{-\pi}{L_xL_y}\sum_{i,j}q_iq_j\left[ z_{ij} {\rm erf}(\alpha z_{ij}) +\frac{1}{\alpha\sqrt\pi}e^{-\alpha^2z_{ij}^2} \right]. \end{equation}
The term ${\mathcal U}_F^0$ is the limit of ${\mathcal U}_F^h$ as $h$ approaches $0$. The singularity of the reciprocal space sum has been removed under the
condition of electroneutrality. An alternative expression of ${\mathcal U}_F^h$ is written as an integral form:
\begin{equation} \label{eq:uewald2dFha}
%\textcolor{black}{{\mathcal U}_F^h}
{\mathcal U}_F^h=\frac{1}{L_xL_y} \sum_{{\mathbf h}\neq 0 }
\int_{-\infty}^{\infty}du\, \frac{ e^{-\frac{ h^2+u^2}{ 4\alpha^2} }}{h^2+u^2 }
 \left| \sum_{j=1}^N q_je^{i{\mathbf h}\cdot{{\mathbf r}_j}} e^{iuz_j} \right|^2 .
\end{equation}
Eqs.~\eqref{eq:uewald2dR} to ~\eqref{eq:uewald2dFha} are the usual Ewald2D formulas. These expressions have been derived in many different ways by
Parry\cite{Parry1975}, by Heyes, Barber and Clarke\cite{Heyes_Clarke1977}, by de Leeuw and Perram\cite{DeLeeuw_Perram1979}, by Grzybowski, Gwozdz and
Brodka\cite{Grzybowski_Brodka2000}, by Minary {\it et. al.}\cite{Tuckerman_Martyna2002} and very recently by Lindbo and Tornberg\cite{Lindbo_Tornberg2012}. 
Derivation of the above equations based on knowledge of elementary calculus is doable but we are not going to show this in the content of current work as excellent
derivations have been done many times in the past. Instead, we emphasize that unlike the case of 3D periodicity, Ewald2D expressions in eqs.~\eqref{eq:uewald2dR} to
~\eqref{eq:uewald2dFha} are exact subject to neutrality condition only irrespective of the shape of the system boundary at infinity. For a neutral system with 2D
periodicity, the convergence of the series in eq.~\eqref{eq:uorigin} does not depend on how the 2D summation vector ${\mathbf n}$ approaches infinity. On the contrast,
for a neutral system with 3D periodicity, the value of the series in eq.~\eqref{eq:uorigin} depends on the behavior that the 3D summation vector approaches
infinity.

Calculation of the Fourier part sum using eqs.~\eqref{eq:uewald2dFh} requires evaluation of pairs $i,j$ and is thus of  ${\cal O}(N^2)$ complexity. Efficient
methods starting from eq.~\eqref{eq:uewald2dFha} have been developed to introduce significant reduction of computational cost
\cite{Kawata_Mikami2001,Lindbo_Tornberg2012}. %As the term ${\mathcal U}_F^0$ in eq.~\eqref{eq:uewald2dF0} only depends on $z$, previous developments have used an
%interpolation method based on B-splines functions\cite{Kawata_Mikami2001} or Chebyshev polynomials\cite{Lindbo_Tornberg2012} which is purely numerical irrespective
%of the intrinsic mathematical function in eq.~\eqref{eq:uewald2dF0}. In the next section, we will show that how one can rewrite the expression for ${\mathcal U}_F^0$
%to achieve a natural expansion which requires only computational cost of ${\cal O}(N)$. 
Alternative expression at the computational cost of ${\cal O}(N)$ using a power-series expansion for the term ${\mathcal U}_F^0$ has been given by 
Minary {\it et. al.} \cite{Tuckerman_Martyna2002}. In the next section, we will show that how one can rewrite the expression for ${\mathcal U}_F^0$ to achieve a
natural expansion which requires only computational cost of ${\cal O}(N)$ when the system is extended along the periodic  $x$ and $y$ directions.

Although eq.(8) carries a computational cost of ${\cal O}(N)$ for the sum over the total number of the charges, the computational cost in total from a
straightforward implementation of eq.(8) is not rigorously ${\cal O}(N)$. In a typical case of extending the system in the periodic $x$ and $y$ directions, the
computation cost for the sum over the vector ${\mathbf h}$ grows as the product of $L_x$ and $L_y$ and thus of ${\cal O}(N)$ itself at a required accuracy.
 The overall computational complexity of eq.(8) is thus ${\cal O}(N^2)$. Excellent developments analog to the well established 3D PME method using grid interpolation and
FFT treatment have been done recently by Lindbo and Tornberg\cite{Lindbo_Tornberg2012}. The current work will not aim at providing a competitive efficient algorithm
for the ${\cal U}_F^h$ term but to analyze analytically the error bounds when approximating the Fourier integral. As discussed previously\cite{Lindbo_Tornberg2012},
implementation of Particle Mesh techniques to the Fourier integral part of the Ewald2D expression will necessarily involve an approximation to the Fourier integral.
Our derivation of the error bounds will show that how the up limit of the accuracy is determined by the combination of appropriate parameters. 
%%%%%%%%%%%%%%%%%%%%%%%%%%%%%%%%%%%%%%%%%%%%%%%%%%%%%%%%%%%%%%%%%%%%%%%%%%%%%%%%%%%%%%%%%%%%%%%%%%%%%%%%%%%%%%%%%%%%%%%%%%%%%%%%%%%%%%%%%%%%%%%%%%%%%%%%%%%%%%%
\section{Alternative expression of ${\mathcal U}_F^0$} \label{sec:alternative}
The equivalence between the two expressions for the term ${\mathcal U}_F^h$ in Eq.~\eqref{eq:uewald2dFha} and Eq.~\eqref{eq:uewald2dFh} was known from the
mathematical identity (see mathematical book\cite{Erdelyi1954}) as early as in 1970s\cite{Heyes_Clarke1977,DeLeeuw_Perram1979}:
% and also a proof of this from elementary calculus in the section~\ref{subsec:apEwald2D}):
\begin{eqnarray} \label{eq:equalFh}
I^h(\omega,\nu) &=& \frac{\pi}{2\omega}\left[ e^{\omega\nu}{\rm erfc}(\omega + \frac{\nu}{2}) + e^{-\omega\nu} {\rm erfc}(\omega - \frac{\nu}{2} ) \right] \nonumber \\
              &=& e^{-\omega^2} \int_{-\infty}^{\infty} dt\,  \frac{e^{-t^2}}{\omega^2 + t^2} e^{it\nu}, 
\end{eqnarray}
where for simplicity, we have used dimensionless quantity $t = u/(2\alpha)$, $\omega = h/(2\alpha) $, and $\nu = 2\alpha z_{ij}$. 
In line with the limit of $\omega\to 0$ in the above Eq. ~\eqref{eq:equalFh} with the divergence removed, we are able to write the following mathematical identity:
\begin{eqnarray} \label{eq:equalF0}
I^0(\nu) &=&  \pi\left[ \nu {\rm erfc}\left(\frac{\nu}{2}\right) - \nu \right] -2\sqrt{\pi} e^{-\nu^2/4} \nonumber \\
         &=&   \int_{-\infty}^{+\infty}dt\, \frac{\displaystyle e^{-t^2}e^{it\nu}-1} {\displaystyle t^2}. 
\end{eqnarray}
When numerically evaluating the integral, we often apply the usual trapezoidal approximation to the integral with infinity in upper and lower limits:
%\begin{eqnarray} \label{eq:trape} I \equiv \int_{-\infty}^{\infty} dt g(t) \, \simeq \sum_{m=-\infty}^{\infty} \zeta g(m\zeta)   
%= \zeta g(0)+\sum_{m=1}^{\infty} \zeta\left[  g(m\zeta)+ g(-m\zeta)\right] \equiv S(\zeta), \end{eqnarray}
\begin{multline} \label{eq:trape} I \equiv \int_{-\infty}^{\infty} dt g(t) \, \simeq \sum_{m=-\infty}^{\infty} \zeta g(m\zeta)   \\
= \zeta g(0)+\sum_{m=1}^{\infty} \zeta\left[  g(m\zeta)+ g(-m\zeta)\right] \equiv S(\zeta), \end{multline}
where an extra parameter of $\zeta$ is used as a small mesh size. Therefore, the right hand side of Eq.~\eqref{eq:equalF0} can be evaluated approximately as:
%\begin{eqnarray} \label{eq:trapeI0} I^0(\nu)\simeq S^0(\nu,\zeta) = -\zeta(1+\frac{\nu^2}{2}) +  
%2\zeta\sum_{m=1}\frac{\displaystyle e^{-(m\zeta)^2}}{ (m\zeta)^2 } \cos( m\nu\zeta)- 2\zeta \sum_{m=1}\frac{1}{(m\zeta)^2}. \end{eqnarray}
\begin{multline} \label{eq:trapeI0} I^0(\nu)\simeq S^0(\nu,\zeta) = -\zeta(1+\frac{\nu^2}{2}) +  \\ 
2\zeta\sum_{m=1}\frac{\displaystyle e^{-(m\zeta)^2}}{ (m\zeta)^2 } \cos( m\nu\zeta)- 2\zeta \sum_{m=1}\frac{1}{(m\zeta)^2}. \end{multline}
Note that we have computed the limit value for the $m=0$ term of the series in eq.~\eqref{eq:trape}:
\begin{equation}\label{eq:0term}  \lim_{m\to 0} \frac{ e^{-(m\zeta)^2}e^{im\zeta\nu}-1 }{(m\zeta)^2} = - (1+ \frac{\nu^2}{2}). \end{equation}
Substituting $\nu = 2\alpha z_{ij}$ in eqs.~\eqref{eq:equalF0},~\eqref{eq:trape}, and~\eqref{eq:trapeI0} and realizing the fact that 
$\sum_{i,j} q_i q_j z_{ij}^2= -2 (\sum_j q_j z_j)^2 $ subject to the electroneutrality condition $\sum_j q_j = 0$,
The formula of ${\mathcal U}_F^0$ in eq.~\eqref{eq:uewald2dF0} can now be rewritten as a form with the computational cost of ${\mathcal O}(N)$:
%\begin{eqnarray} \label{eq:uewald2dF0a}
%%\textcolor{black}{{\mathcal U}_F^0}
%{\mathcal U}_F^0\simeq \frac{2\zeta\alpha}{ L_xL_y} \left(\sum_j q_j z_j \right)^2 + 
%             \frac{\zeta}{\alpha L_xL_y} \sum_{m=1} \frac{ e^{-(m\zeta)^2}}{ (m\zeta)^2}  \left|\sum_j q_j e^{i2m\alpha z_j\zeta} \right|^2,
%\end{eqnarray}
\begin{multline} \label{eq:uewald2dF0a}
%\textcolor{blue}{{\mathcal U}_F^0}
{\mathcal U}_F^0\simeq \frac{2\zeta\alpha}{ L_xL_y} \left(\sum_j q_j z_j \right)^2 + \\
             \frac{\zeta}{\alpha L_xL_y} \sum_{m=1} \frac{ e^{-(m\zeta)^2}}{ (m\zeta)^2}  \left|\sum_j q_j e^{i2m\alpha z_j\zeta} \right|^2,
\end{multline}
where the approximation follows from the use of trapezoidal sum in eq.~\eqref{eq:trapeI0}. Evaluation of the term ${\cal U}_F^0$ from the above
eq.~\eqref{eq:uewald2dF0a} is a natural expansion of the Fourier integral to the charge density in the reciprocal space and might be superior to the previous development
based on interpolation methods which do not contain any intrinsic information from the Fourier transform.
%%%%%%%%%%%%%%%%%%%%%%%%%%%%%%%%%%%%%%%%%%%%%%%%%%%%%%%%%%%%%%%%%%%%%%%%%%%%%%%%%%%%%%%%%%%%%%%%%%%%%%%%%%%%%%%%%%%%%%%%%%%%%%%%%%%%%%%%%%%%%%%%%%%%%%%%%%%%%%%%
\section{Error bounds for the trapezoidal approximation} \label{sec:error}
\begin{figure}
\centerline{
\psfig{file=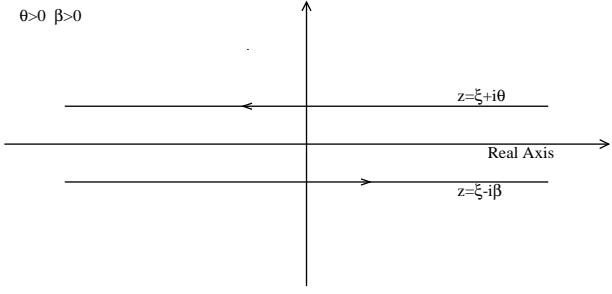,width=3.2in,angle=270}}
\caption{Illustration of the path when there is no pole in the complex plane for the integrand. Integration along the two straight lines gives the error term.
See eq.~\eqref{eq:moriI0} and~\eqref{eq:errorE0}.}
\label{fig:contournopole} \end{figure}
Using the trapezoidal sum to approximately evaluate the Fourier integrals $U_F^0$ and $U_F^h$ both generate errors. In general, when the chosen mesh size $\zeta$ is
small enough, excellent accuracy can be achieved given that enough terms in the trapezoidal series are computed. However, one might always want to save computational
cost by using relatively larger value of $\zeta$ at a required accuracy of computation. It is still unknown that how the accuracy of eq.~\eqref{eq:uewald2dF0a} is
controlled by the choice of parameter $\zeta$. We now proceed to discuss in general how the setup of the mesh size $\zeta$ affects the error due to the use of the
trapezoidal approximation $S(\zeta)$ to the integral $I$ in eq.~\eqref{eq:trape}. We start from a rigorous mathematic formula connecting an arbitrary integral to its
trapezoidal sum which states that the definite integral over infinity in real axis can be expressed rigorously as a combination of the trapezoidal sum, a correction
term and an error term:
\begin{equation} \label{eq:mori}
 I = \int_{-\infty}^{\infty} dt\, g(t)  = S(\zeta) + C(\zeta) + E(\zeta), \end{equation}
where $S(\zeta)$ is the trapezoidal sum as in eq.~\eqref{eq:trape}. The error term $E(\zeta)$ is expressed as an integration along a path in the complex plane:
\begin{equation} \label{eq:errorC}
E(\zeta) =   \frac{1}{2\pi i} \oint_{\rm path} dz\, \Psi(z)g(z) ,
\end{equation}
where $\Psi(z)$ is the characteristic function in the complex plane defined as:
\begin{equation}\label{eq:char} \Psi(z) = \frac{\displaystyle \mp 2\pi i}{\displaystyle 1-e^{\mp 2\pi i z/\zeta} } \mbox{\quad\quad} {\rm Im}(z)\gtrless 0 .
\end{equation}
The characteristic function $\Psi(z)$ depends on the mesh size parameter $\zeta$ but does not depend on the form of the integrand $g(z)$. Depending on the nature of
the integrand $g(z)$, the correction term $C(\zeta)$ could be zero or proportional to the residue of the complex function $\Psi(z)g(z)$ at the singularity point of $g(z)$.
Eq.~\eqref{eq:errorC} expresses the difference between the integral and the corresponding trapezoidal form to the integral along the certain path in the complex
plane. This powerful equation has been used to develop the double-exponential transform in the field of mathematics\cite{Mori2005}. For the ease of reading
this work, we provide a brief proof of the above important eqs~\eqref{eq:mori} to ~\eqref{eq:char} in Appendix~\ref{subsec:apTrapezoidal}. Note that the symbol
$z$ in these equations stands for the variable in the complex plane which should not be confused with the previous notation of $z$ used as the variable of position or
length. The error term $E(\zeta)$ is the value of the integration along the contour and thus depends on both the form of the
integrand and the choice of the path. When two infinite lines parallel to the real axis are chosen as the contour (see Fig.~\ref{fig:contournopole}) and the
integrand in eq.~\eqref{eq:equalF0} has no pole in the interior formed by the contour, the correction term $C(\zeta)$ is simply zero and the error term $E(\zeta)$ is
the integral along the two straight lines (see details in Appendix~\ref{subsec:apTrapezoidal}). We thus obtain an exact expression for the integral $I^0(\nu)$:
\begin{equation}\label{eq:moriI0}
I^0(\nu) = S^0(\nu,\zeta) + E^0(\nu,\zeta) ,
\end{equation}
where $S^0(\nu,\zeta)$ is the expression in eq.~\eqref{eq:trapeI0} and the error term is written as:
%\begin{eqnarray} \label{eq:errorE0}
%E^0(\nu,\zeta) =  \frac{1}{2\pi i}\int_{z=\xi+i\theta}dz\, \Psi(z)g^0(z) 
%+\frac{1}{2\pi i}\int_{z=\xi-i\beta} dz\, \Psi(z)g^0(z), \end{eqnarray}
\begin{multline} \label{eq:errorE0}
E^0(\nu,\zeta) =  \frac{1}{2\pi i}\int_{z=\xi+i\theta}dz\, \Psi(z)g^0(z) \\
+\frac{1}{2\pi i}\int_{z=\xi-i\beta} dz\, \Psi(z)g^0(z), \end{multline}
and 
\begin{equation} \label{eq:g0z}
g^0(z) = \frac{ e^{-z^2} e^{iz\nu} - 1  }{ z^2 } .
\end{equation}
We now directly manipulate the complex integral such that an appropriate error bound $\delta(\nu,\zeta)$ exists:
\begin{equation}\label{eq:ebI0} \left|  E^0(\nu,\zeta)  \right| \leqslant \delta(\nu,\zeta) . \end{equation}
\begin{figure}[udtp]
\centerline{
\psfig{file=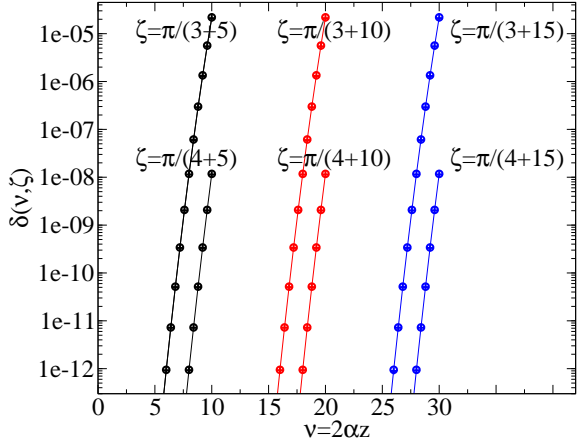,width=3.0in,angle=270} }
\caption{Analytical error bounds ($\delta(\nu,\zeta)$: $\circ$) compared to the computed error ($E^0(\nu,\zeta)$: $+$) at different values of $\zeta$. The
exact value of $I^0(\nu)$ is calculated from the expression with complementary error function in eq.~\eqref{eq:equalF0} and $E^0(\nu,\zeta)$ is thus evaluated as
$E^0(\nu,\zeta)=I^0(\nu)-S^0(\nu,\zeta)$ (see eq.~\eqref{eq:moriI0}). The plus symbols overlap with the circle symbols indicating that the evaluation of error bounds using
eq.~\eqref{eq:exactebI0} is almost exact. Note that when comparing the error and its error bound, we have removed a constant accounting for the difference at $\nu=0$.}
\label{fig:errorboundsz}
\end{figure}
Because one can always numerically determine the difference at $\nu=0$ between the trapezoidal sum $S^0(\nu,\zeta)$ and $I^0(\nu)$ evaluated as in eq.~\eqref{eq:equalF0},
it is much simpler but enough to evaluate the integration over the major part of $g^0(z)$ which is:
\begin{equation} \label{eq:g0mz}
g^0_m(z) = \frac{ e^{-z^2} e^{iz\nu}}{z^2} .
\end{equation}
Each of the integration along the two straight lines in Fig.\ref{fig:contournopole} has its own up limit:
%\begin{eqnarray}\label{eq:e0bu} \left|  \frac{1}{2\pi i} \int_{z=\xi+i\theta} dz\, \Psi(z)g^0_m(z) \right| < \delta^u(\nu,\zeta,\theta) 
% = \frac{\displaystyle \sqrt{\pi} e^{\left[\theta-(\pi/\zeta+\nu/2)\right]^2} }{\displaystyle \theta^2(1-e^{-2\pi\theta/\zeta}) }e^{-(\pi/\zeta + \nu/2)^2} ,
%\end{eqnarray}
%and 
%\begin{eqnarray}\label{eq:e0bd} \left|  \frac{1}{2\pi i} \int_{z=\xi-i\beta} dz\, \Psi(z)g^0_m(z) \right| < \delta^d(\nu,\zeta,\beta)  
%= \frac{\displaystyle \sqrt{\pi} e^{\left[\beta-(\pi/\zeta-\nu/2)\right]^2} }{\displaystyle \beta^2(1-e^{-2\pi\beta/\zeta}) }e^{-(\pi/\zeta - \nu/2)^2}.
%\end{eqnarray}
\begin{multline}\label{eq:e0bu} \left|  \frac{1}{2\pi i} \int_{z=\xi+i\theta} dz\, \Psi(z)g^0_m(z) \right| < \delta^u(\nu,\zeta,\theta) \\
 = \frac{\displaystyle \sqrt{\pi} e^{\left[\theta-(\pi/\zeta+\nu/2)\right]^2} }{\displaystyle \theta^2(1-e^{-2\pi\theta/\zeta}) }e^{-(\pi/\zeta + \nu/2)^2} ,
\end{multline}
and
\begin{multline}\label{eq:e0bd} \left|  \frac{1}{2\pi i} \int_{z=\xi-i\beta} dz\, \Psi(z)g^0_m(z) \right| < \delta^d(\nu,\zeta,\beta)  \\
= \frac{\displaystyle \sqrt{\pi} e^{\left[\beta-(\pi/\zeta-\nu/2)\right]^2} }{\displaystyle \beta^2(1-e^{-2\pi\beta/\zeta}) }e^{-(\pi/\zeta - \nu/2)^2}.
\end{multline}
The right hand sides of the above two equations have minimum values at $\theta_0\simeq \pi/\zeta+\nu/2$ and $\beta_0\simeq \pi/\zeta-\nu/2$ respectively. The total error
bound for the evaluation of $I^0(\nu)$ using the trapezoidal approximation is simply the sum of the above two error bounds at their minima:
\begin{equation}\label{eq:exactebI0} \delta(\nu,\zeta) = \delta^u(\nu,\zeta,\theta_0)  + \delta^d(\nu,\zeta,\beta_0).   \end{equation}
A proof of eqs.~\eqref{eq:e0bu} and~\eqref{eq:e0bd} can be found in Appendix~\ref{subsec:errorbound}.
Fig.~\ref{fig:errorboundsz}  shows direct calculations of the error (plus symbols) introduced when using trapezoidal approximation in eq.~\eqref{eq:trapeI0} to the
integral $I^0(\nu)$ in eq.~\eqref{eq:equalF0} as well as the total error bound (circle symbols) as a function of chosen parameter $\nu$ for largest $\nu$ values
taken to be $10$, $20$ and $30$ respectively. The overlap between the errors and the error bounds shown in Fig.~\ref{fig:errorboundsz} indicates that using
Eq.~\eqref{eq:exactebI0} for the error bound is pretty rigorous. Because the error bound is a Gaussian function of $\pi/\zeta - \nu/2$, strong error control can be
achieved by simply adjusting the value of $\zeta$ when the system length scale $z$ becomes large. When plotting in the logarithm scale, the Gaussian form of the error
bounds behaves linearly as in Fig.~\ref{fig:errorboundsz}.

\begin{figure}
\centerline{
\psfig{file=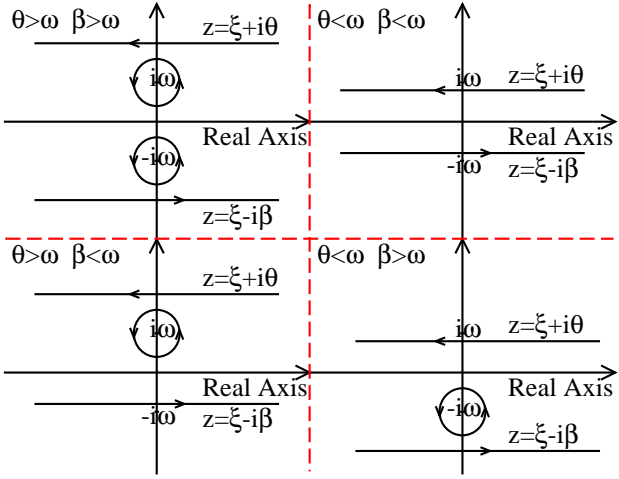,width=3.2in,angle=270}}
\caption{Illustration of the path when there are two simple poles ($\pm i\omega$) for the integrand $g^h(z)$. The expression of the correction term $C^h(\omega,\nu,\zeta)$ depends on the
relative position of $\omega$ to $\theta$ or $\beta$. When the pole is inside the interior formed by the two straight lines, the corresponding integral around the pole
(residue) is then included into the correction term. When the pole is outside the region, there is no need to perform the integral along the contour ($C_+$ or
$C_-$).}
\label{fig:contourpoles}
\end{figure}

For the case of more complicated integral $I^h(\omega,\nu)$ in eq.~\eqref{eq:equalFh}, the integrand in the complex plane has two symmetric first order poles (simple
poles) at points $z=\pm i\omega$:
\begin{equation} \label{eq:ghz}
g^h(z) = \frac{ e^{-(\omega^2+z^2)} e^{iz\nu} }{\omega^2+z^2}.  \end{equation}
Now we proceed to see how we can choose appropriate contour lines such that the error term is minimized. Similar to the case of $I^0(\nu)$, the integral in
eq.~\eqref{eq:equalFh}  can be written as (see details in Appendix~\ref{subsec:errorbound}). 
\begin{equation} \label{eq:moriIh}
I^h(\omega,\nu) \equiv S^h(\omega,\nu,\zeta) + C^h(\omega,\nu,\zeta) +  E^h(\omega,\nu,\zeta),
\end{equation}
where the trapezoidal sum is
\begin{eqnarray}
S^h(\omega,\nu,\zeta)&=&\zeta e^{-\omega^2}\left[ \frac{1}{\omega^2} +\sum_{m=1}^{\infty}\frac{2 e^{-(m\zeta)^2} \cos(m\zeta\nu) }{\omega^2 + (m\zeta)^2}\right] 
 \nonumber \\        &=& \zeta e^{-\omega^2} \sum_{m=-\infty}^{\infty} \frac{e^{-(m\zeta)^2} e^{im\zeta\nu }}{\omega^2 + (m\zeta)^2 },
 \label{eq:sh} \end{eqnarray}
and the error term is an integration along the two straight lines shown in Fig.~\ref{fig:contourpoles}:
%\begin{eqnarray}\label{eq:eh}
%E^h(\omega,\nu,\zeta) = \frac{1}{2\pi i}\int_{z=\xi+i\theta}dz\, \Psi(z)g^h(z) 
%                      + \frac{1}{2\pi i}\int_{z=\xi-i\beta} dz\, \Psi(z)g^h(z) .
%\end{eqnarray}
\begin{multline}\label{eq:eh}
E^h(\omega,\nu,\zeta) = \frac{1}{2\pi i}\int_{z=\xi+i\theta}dz\, \Psi(z)g^h(z) \\
                      + \frac{1}{2\pi i}\int_{z=\xi-i\beta} dz\, \Psi(z)g^h(z) .
\end{multline}
The correction term is related to the negative residue of the pole in the complex plane. According to residue theorem, we have
\begin{equation}\label{eq:residueu}
\frac{-1}{2\pi i}\oint_{C_+}dz\, \Psi(z)g^h(z) = \frac{\pi}{\omega} \frac{e^{-\omega\nu}}{1-e^{2\pi\omega/\zeta}} ,
\end{equation}
and 
\begin{equation}\label{eq:residued}
\frac{-1}{2\pi i}\oint_{C_-}dz\, \Psi(z)g^h(z) = \frac{\pi}{\omega} \frac{e^{\omega\nu}}{1-e^{2\pi\omega/\zeta}} .
\end{equation}
When either pole is inside the interior formed by the two straight lines, its corresponding residue will be included in the correction term $C^h(\omega,\nu,\zeta)$.
Otherwise, the pole makes no contribution to the correction term. For example, in the case of $\theta > \omega$ and $\beta > \omega $, both poles are inside the
region (see the top-left part of Fig.~\ref{fig:contourpoles}). A rigorous expression for the correction term is written as:
\begin{equation}\label{eq:ch}
C^h(\omega,\nu,\zeta)= \frac{\pi}{\omega} \frac{e^{-\omega\nu}}{1-e^{2\pi\omega/\zeta}} + \frac{\pi}{\omega} \frac{e^{\omega\nu}}{1-e^{2\pi\omega/\zeta}} .
\end{equation}
When both poles are outside the two straight lines ($\theta < \omega  $ and $ \beta < \omega $, see the top-right part of Fig.~\ref{fig:contourpoles}),
 the correction term $C^h(\omega,\nu,\zeta)$ is thus simply zero. 

\begin{figure}[udtp]
\centerline{
\psfig{file=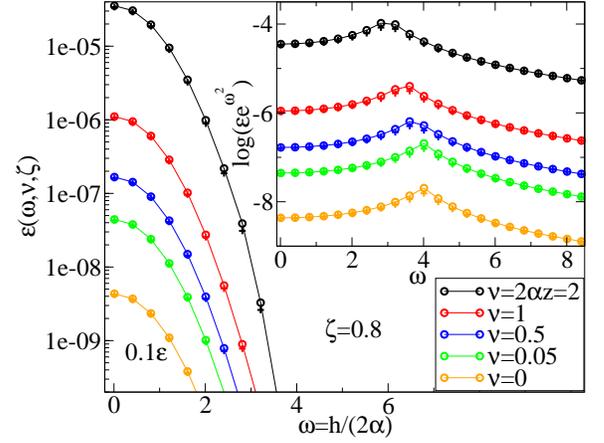,width=3.0in,angle=270}}
\caption{Error bounds ($\varepsilon(\omega,\nu,\zeta)$:$\circ$) compared to the computed error ($E^h(\omega,\nu,\zeta)$: $+$) as a function of $\omega$
for a given value of $\zeta=0.8$ and a series of values of $\nu$. Insets show the same data but multiplied by a factor of $e^{\omega^2}$. The line for $\nu=0$ is very
closed to the line at $\nu=0.05$ and is thus multiplied by $0.1$ for easy view. See eq.~\eqref{eq:exactebIh} for the evaluation of error bounds and
eqs.~\eqref{eq:equalFh} and~\eqref{eq:moriIh} for the numerical error of using the trapezoidal sum and the correction term to approximate the integral $I^h(\omega,\nu)$.}
\label{fig:errorboundslvp8}
\end{figure}

\begin{figure}[udtp]
\centerline{
\psfig{file=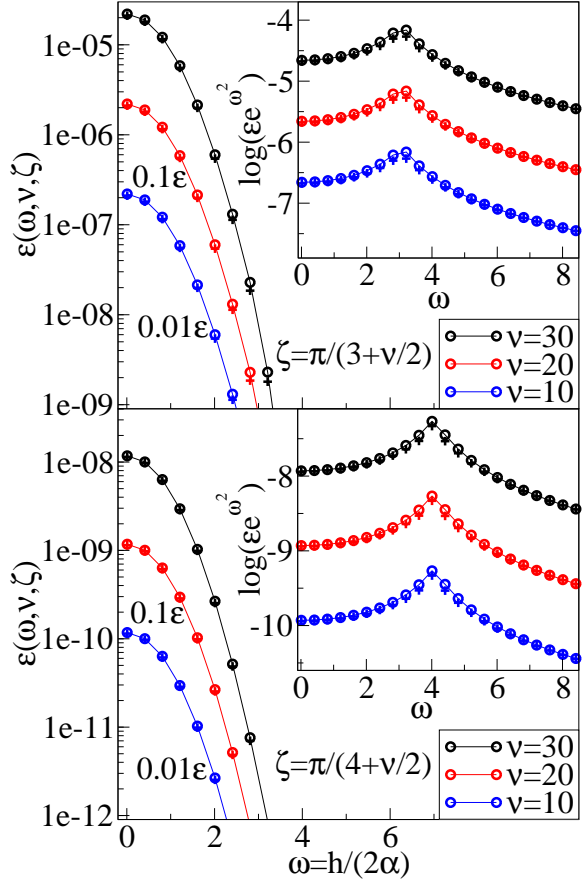,width=3.0in,angle=0}}
\caption{Same as in the previous Fig.~\ref{fig:errorboundslvp8} but different values of $\zeta$: $\pi/\zeta - \nu/2 = 3$ (top)  and $\pi/\zeta - \nu/2 = 4$ (bottom).
Insets show the same data but multiplied by a factor of $e^{\omega^2}$. For easier view, $\epsilon$ values at $\nu=20$ and $\nu=10$ have been multiplied by $0.1$ and
$0.01$ respectively. Be aware of that the error is shrunk by 3 orders of magnitude when $\zeta$ becomes slightly smaller.}
\label{fig:errorboundsd3d4}
\end{figure}

To get the error bound for the error term $E^h(\omega,\nu,\zeta)$ such that:
\begin{equation}\label{eq:ebIh}
\left| E^h(\omega,\nu,\zeta)  \right| \leqslant \varepsilon(\omega,\nu,\zeta),
\end{equation}
we further find that (see details in appendix~\ref{subsec:errorbound}):
%\begin{eqnarray}\label{eq:ehbu}
%\left|  \frac{1}{2\pi i} \int_{z=\xi+i\theta} dz\, \Psi(z)g^h(z) \right| < \varepsilon^u(\omega,\nu,\zeta,\theta) 
%= \frac{\displaystyle e^{-\omega^2} \sqrt{\pi} e^{\left[\theta-(\pi/\zeta+\nu/2)\right]^2} }
%{\displaystyle |\theta^2-\omega^2|(1-e^{-2\pi\theta/\zeta}) }e^{-(\pi/\zeta + \nu/2)^2}, \end{eqnarray}
%and
%\begin{eqnarray}\label{eq:ehbd} 
%\left|  \frac{1}{2\pi i} \int_{z=\xi-i\beta} dz\, \Psi(z)g^h(z) \right| < \varepsilon^d(\omega,\nu,\zeta,\beta) 
%=  \frac{\displaystyle e^{-\omega^2} \sqrt{\pi} e^{\left[\beta-(\pi/\zeta-\nu/2)\right]^2} }
%{\displaystyle |\beta^2-\omega^2|(1-e^{-2\pi\beta/\zeta}) }e^{-(\pi/\zeta - \nu/2)^2}. \end{eqnarray}
\begin{multline}\label{eq:ehbu}
\left|  \frac{1}{2\pi i} \int_{z=\xi+i\theta} dz\, \Psi(z)g^h(z) \right| < \varepsilon^u(\omega,\nu,\zeta,\theta) \\
= \frac{\displaystyle e^{-\omega^2} \sqrt{\pi} e^{\left[\theta-(\pi/\zeta+\nu/2)\right]^2} }
{\displaystyle |\theta^2-\omega^2|(1-e^{-2\pi\theta/\zeta}) }e^{-(\pi/\zeta + \nu/2)^2}, \end{multline}
and
\begin{multline}\label{eq:ehbd} 
\left|  \frac{1}{2\pi i} \int_{z=\xi-i\beta} dz\, \Psi(z)g^h(z) \right| < \varepsilon^d(\omega,\nu,\zeta,\beta) \\
=  \frac{\displaystyle e^{-\omega^2} \sqrt{\pi} e^{\left[\beta-(\pi/\zeta-\nu/2)\right]^2} }
{\displaystyle |\beta^2-\omega^2|(1-e^{-2\pi\beta/\zeta}) }e^{-(\pi/\zeta - \nu/2)^2}. \end{multline}
Assuming $\theta=\theta_0$ and $\beta=\beta_0$ are to minimize the right hand side of eqs.~\eqref{eq:ehbu} and~\eqref{eq:ehbd} respectively, the total error bound is
taken to be the sum of the minima:
\begin{equation}\label{eq:exactebIh}
\varepsilon(\omega,\nu,\zeta) = \varepsilon^u(\omega,\nu,\zeta,\theta_0) + \varepsilon^d(\omega,\nu,\zeta,\beta_0).   \end{equation}
A comparison between $\varepsilon(\omega,\nu,\zeta)$ and $E^h(\omega,\nu,\zeta)$ at given mesh size $\zeta=0.8$ is shown as circle symbols and plus symbols
respectively in Fig.~\ref{fig:errorboundslvp8}.
Clearly, the error bounds are always in the same magnitude as the exact errors indicating that eq.~\eqref{eq:exactebIh} is rigorous when used to control the errors
generated from the trapezoidal approximation.
Because the error bound is proportional to a Gaussian function of $\pi/\zeta - \nu/2$, adjusting the mesh size $\zeta$ will dramatically change the accuracy.
Fig.~\ref{fig:errorboundsd3d4} shows that how the change of $\zeta$ affects the value of the error bound $\varepsilon(\omega,\nu,\zeta)$ with comparison to the exact error
computed numerically.

Clearly for both cases of Fourier integrals ${\mathcal U}_F^h$ and ${\mathcal U}_F^0$, we have shown that numerical analysis of the errors generated from trapezoidal
approximation to $I^h(\omega,\nu)$ and $I^0(\nu)$ supports our rigorous analytical error bounds of Gaussian functions of $\pi/\zeta - \nu/2$.
 ${\mathcal U}_F^h$ and ${\mathcal U}_F^0$ are combinations of pairs of product of charges and thus much more complicated. However, the error control through the
Gaussian decay function should still work well. In a word, we have started from the integral expression eq.~\eqref{eq:uewald2dFha} and analytically proved that the
error bounds generated from the trapezoidal approximation to eq.~\eqref{eq:uewald2dFha} and eq.~\eqref{eq:uewald2dF0} scale as Gaussian functions of the appropriate
combination of system parameters and can thus be well controlled.

\begin{figure}[udtp]
\centerline{
\psfig{file=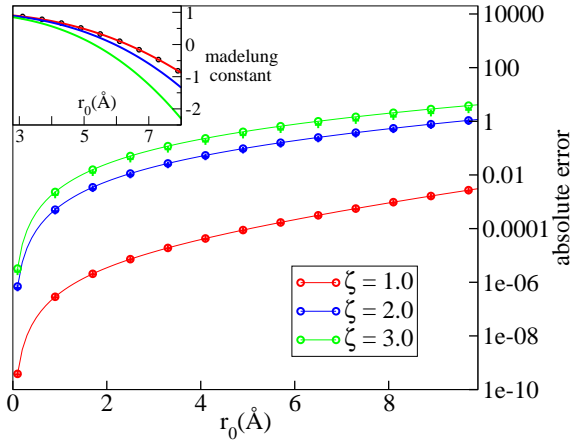,width=3.0in,angle=270}}
\caption{Error bounds ($\circ$) compared to the computed errors ($+$) for the evaluation of Madelung constants in a series of model ionic crystals in slab
geometry. Inset shows the values of Madelung constants computed at the given parameters $\zeta$ with the exact values shown as black circles. The screening factor
$\alpha$ is set to $0.1$\AA$^{-1}$. See eqs.~\eqref{eq:exactebI0}, ~\eqref{eq:exactebIh} and~\eqref{eq:madelung}.}
\label{fig:madelung}
\end{figure}

To check the validity of the present formula of the error bounds in a realistic example, we compute the Madelung constants in model ionic crystals.
The Madelung constant of the $i$-th charge in the slab geometry ($M^s_i$) is defined as the total electrostatic potential felt by the $i$-th particle in the infinite
2D periodic system normalized to the electrostatic potential between the nearest neighbor pair of charges:
\begin{equation} \label{eq:madelung} M^s_i = \left[ \sideset{}{'}\sum_{\mathbf n} \left(\sum_{i=1}^N \sum_{j=1}^N \frac{ \displaystyle q_iq_j }{\displaystyle \left|
 {\mathbf n} + {\mathbf r}_{ij} \right| } \right) \right] \frac{r_0}{q_1q_2} ,
\end{equation}
where $r_0$ is the distance between the nearest neighbor charges $q_1$ and $q_2$. The prime indicates that the $i=j$ term is omitted in case of $n_x=n_y=0$.
Notations in the above equation are the same as in eq.~\eqref{eq:uorigin}. %
We set up a simple toy model consisted of one cation and one anion both with unit charge and located at the $z$ axis. The unit cell lengths in $x$ and $y$
direction are chosen as $L_x=L_y=10$ \AA. Fig.~\ref{fig:madelung} compares the computed error and the error bounds as well as the evaluation of the Madelung constant as a function of the lattice constant
($r_0$). Clearly, the analytical formula of the error bounds work for an arbitrary range of accuracy control (up to machine accuracy) for the computation of Madelung constants of the model 
crystals.

%%%%%%%%%%%%%%%%%%%   Boundary correction and electrostatic Layer correction terms  %%%%%%%%%%%%%%%%%%%%%%%%%%%%%%%%%%%%%%%%%%%%%%%%%%%%%%%%%%%%%%%%%%%%%%%%%%%%%%%%
\section{The boundary correction and electrostatic layer correction terms}\label{sec:terms}

The above derivation followed the Fourier integral expressions and then employed trapezoidal approximation. Alternative methodologies relating the 2D electrostatics to
the well known Ewald expression for 3D periodicity (Ewald3D) have been developed and widely used\cite{Yeh_Berkowitz1999,Arnold_Holm2002}. In their development, the
nonperiodic $z$ dimension is extended to periodicity with a chosen unit length of $L_z$ (see Fig.~\ref{fig:period2D}). To correct the effect of modified boundary
condition and added extra layers in the $z$ direction, two terms are added to the usual Ewald3D expression. One is the boundary correction term (BC) in Ewald3Dc
method\cite{Yeh_Berkowitz1999} and the other is the electrostatic layer correction (ELC) term\cite{Arnold_Holm2002b} (see Fig.~\ref{fig:period2D}). Using our notation,
the BC and ELC terms (see eq. 10 of ref.\cite{Yeh_Berkowitz1999} and eq. 11 of ref.\cite{Arnold_Holm2002b}) are written as :
\begin{equation}\label{eq:eqBC} {\cal U}_{\rm BC} = \frac{2\pi}{L_xL_yL_z}\left| \sum_{j=1}^N q_j z_j\right|^2 , \end{equation}
and 
\begin{equation}\label{eq:eqELC} {\cal U}_{\rm ELC} =\frac{2\pi}{L_xL_y} \sum_{i,j}q_iq_j\sum_{h\neq 0} \frac{e^{i{\mathbf h}\cdot {\mathbf r}_{ij}}}{h}
                                                                                                        \frac{\cosh(hz_{ij})}{1-e^{hL_z}} , 
\end{equation}
respectively. We now proceed to see how our rigorous development is related to the previous physical intuitions. Combining
eqs.~\eqref{eq:uewald2dFha},~\eqref{eq:equalFh},~\eqref{eq:moriIh},~\eqref{eq:sh},and~\eqref{eq:ch}, the major part including the trapezoidal term and the correction
terms of the Fourier integral ${\cal U}_F^h$ is written as:
%\begin{eqnarray}
%{\cal U}_F^h & =      &  \frac{1}{2\alpha L_x L_y} \sum_{i,j} q_i q_j \sum_{h\neq0} e^{i{\mathbf h}\cdot{\mathbf r}_{ij}} I^h(\omega,\nu) \nonumber \\
%             & \simeq & \frac{1}{2\alpha L_x L_y} \sum_{i,j} q_i q_j \sum_{h\neq0} e^{i{\mathbf h}\cdot{\mathbf r}_{ij}}\cdot \nonumber 
%             ( S^h(\omega,\nu,\zeta)+C^h(\omega,\nu,\zeta) ) \nonumber \\
%             & =      & \frac{\zeta}{2\alpha L_x L_y} \sum_{m}\sum_{h\neq0} \frac{ e^{-(m\zeta)^2} e^{-h^2/(4\alpha^2)} }{(m\zeta)^2 + h^2/(4\alpha^2) } \cdot 
%\nonumber    \left|\sum_j q_j e^{i({\mathbf h}\cdot{\mathbf r}_j + 2m\alpha\zeta z_j )}  \right|^2  \nonumber \\
%             &        &   + \frac{2\pi}{L_xL_y} \sum_{i,j}\sum_{h\neq0}\frac{q_iq_j}{h} \frac{\cosh(hz_{ij})}{1- e^{\pi h/(\alpha\zeta)} }
%                                          e^{i{\mathbf h}\cdot {\mathbf r}_{ij}} ,
%\label{eq:muFha}\end{eqnarray}
\begin{eqnarray}
{\cal U}_F^h & =      &  \frac{1}{2\alpha L_x L_y} \sum_{i,j} q_i q_j \sum_{h\neq0} e^{i{\mathbf h}\cdot{\mathbf r}_{ij}} I^h(\omega,\nu) \nonumber \\
             & \simeq & \frac{1}{2\alpha L_x L_y} \sum_{i,j} q_i q_j \sum_{h\neq0} e^{i{\mathbf h}\cdot{\mathbf r}_{ij}}\cdot \nonumber \\
             &        & \quad\quad\quad                                                       ( S^h(\omega,\nu,\zeta)+C^h(\omega,\nu,\zeta) ) \nonumber \\
             & =      & \frac{\zeta}{2\alpha L_x L_y} \sum_{m}\sum_{h\neq0} \frac{ e^{-(m\zeta)^2} e^{-h^2/(4\alpha^2)} }{(m\zeta)^2 + h^2/(4\alpha^2) } \cdot 
\nonumber \\ &        & \quad\quad\quad   \left|\sum_j q_j e^{i({\mathbf h}\cdot{\mathbf r}_j + 2m\alpha\zeta z_j )}  \right|^2  \nonumber \\
             &        &   + \frac{2\pi}{L_xL_y} \sum_{i,j}\sum_{h\neq0}\frac{q_iq_j}{h} \frac{\cosh(hz_{ij})}{1- e^{\pi h/(\alpha\zeta)} }
                                          e^{i{\mathbf h}\cdot {\mathbf r}_{ij}} ,
\label{eq:muFha}\end{eqnarray}
where we have considered the correction term corresponding to the top-left part of Fig.~\ref{fig:contourpoles}. Similarly, we have written the major part of the
Fourier integral ${\cal U}_F^0$ as in eq.~\eqref{eq:uewald2dF0a}:
\begin{multline}\label{eq:muF0a}  % this equation is the same as in eq.~\eqref{eq:uewald2dF0a}
{\cal U}_F^0 \simeq \frac{2\zeta\alpha}{ L_xL_y} \left(\sum_j q_j z_j \right)^2 + \\
             \frac{\zeta}{\alpha L_xL_y} \sum_{m=1} \frac{ e^{-(m\zeta)^2}}{ (m\zeta)^2}  \left|\sum_j q_j e^{i2m\alpha z_j\zeta} \right|^2 .
\end{multline}
%
%\begin{eqnarray}\label{eq:muF0a}  % this equation is the same as in eq.~\eqref{eq:uewald2dF0a}
%{\cal U}_F^0 \simeq \frac{2\zeta\alpha}{ L_xL_y} \left(\sum_j q_j z_j \right)^2 + 
%             \frac{\zeta}{\alpha L_xL_y} \sum_{m=1} \frac{ e^{-(m\zeta)^2}}{ (m\zeta)^2}  \left|\sum_j q_j e^{i2m\alpha z_j\zeta} \right|^2 .
%\end{eqnarray}
The error introduced by the trapezoidal approximation has been discussed as combinations of Gaussian functions. Let's set the dimensionless mesh size
$\zeta=\pi/(\alpha L_z)$ and define a Fourier space vector ${\mathbf k} = 2\pi(h_x/L_x, h_y/L_y, m/L_z)$, we have
%\begin{multline}\label{eq:uewald2Dappx}
%{\cal U}_F^h+{\cal U}_F^0\simeq\frac{2\pi}{L_xL_yL_z}\sum_{k\neq0}\frac{e^{-k^2/(4\alpha^2)}}{k^2}\left|\sum_j q_je^{i {\mathbf k}\cdot {\mathbf r}_j}\right|^2 
%          + \frac{2\pi}{L_xL_yL_z} \left( \sum_j q_j z_j \right)^2  \quad\quad\quad\quad\quad\quad \\
%          + \frac{2\pi}{L_xL_y} \sum_{i,j}q_iq_j \sum_{h\neq0} \left[ \frac{e^{i{\mathbf h}\cdot {\mathbf r}_{ij} }}{h}\frac{\cosh(hz_{ij})}{1-e^{hL_z}}\right],
%\end{multline}
\begin{multline}\label{eq:uewald2Dappx}
{\cal U}_F^h+{\cal U}_F^0\simeq\frac{2\pi}{L_xL_yL_z}\sum_{k\neq0}\frac{e^{-k^2/(4\alpha^2)}}{k^2}\left|\sum_j q_je^{i {\mathbf k}\cdot {\mathbf r}_j}\right|^2 \\
          + \frac{2\pi}{L_xL_yL_z} \left( \sum_j q_j z_j \right)^2  \quad\quad\quad\quad\quad\quad \\
          + \frac{2\pi}{L_xL_y} \sum_{i,j}q_iq_j \sum_{h\neq0} \left[ \frac{e^{i{\mathbf h}\cdot {\mathbf r}_{ij} }}{h}\frac{\cosh(hz_{ij})}{1-e^{hL_z}}\right],
\end{multline}
where the first, second, and third term in the right hand side of eq.~\eqref{eq:uewald2Dappx} are the Fourier part of the regular Ewald3D expression, the BC term
(see eq.~\eqref{eq:eqBC}) and the ELC term (see eq.~\eqref{eq:eqELC}) respectively. Therefore, in connection with the previous developments, our analytical formulation
rigorously shows that both ELC\cite{Arnold_Holm2002b} and BC\cite{Yeh_Berkowitz1999} terms
naturally arise from the trapezoidal expressions to the Fourier integral terms ${\mathcal U}_F^h$ and ${\mathcal U}_F^0$ respectively. 
%This derivation further clarifies that the two ways of reducing computational cost are almost identical in term of computational accuracy. 
%However, in the case that $\omega$ is very large and $\zeta$ is small, contribution from the two residues is on the same magnitude as or much smaller than the
%error term $E^h(\omega,\nu,\zeta)$. 
%%%%%%%%%%%%%%%%%%%%%%%%%%%%%%%%%%%%%%% Conclusion %%%%%%%%%%%%%%%%%%%%%%%%%%%%%%%%%%%%%%%%%%%%%%%%%%%%%%%%%%%%%%%%%%%%%%%%%%%%%%%%%%%%%%%%%%%%%%%%%%%%%%%%%%%%%%%%
\section{Conclusion}\label{sec:conclusion}
We have systematically described how the error of using trapezoidal approximation to the Fourier integrals in Ewald2D expression can be controlled through Gaussian
functions of the appropriate parameters. Our analytical derivation illustrated the intrinsic relation between the previous developed Ewald3DC/EwaldELC methods and the
methodologies of using trapezoidal sum to Fourier integrals. The formulation of singularity term in the Fourier integral as a natural expansion to charge densities
in the reciprocal space might be useful to replace the previous interpolation methods.
%%%%%%%%%%%%%%%%%%%%%%%%%%%%%%%%%%%      Appendix   A %%%%%%%%%%%%%%%%%%%%%%%%%%%%%%%%%%%%%%%%%%%%%%%%%%%%%%%%%%%%%%%%%%%%%%%%%%%%%%%%%%%%%%%%%%%%%%%%%%%%%%%%%%%%%
\section{Appendix}
\subsection{Exact correction to the trapezoidal approximation of an integral}
\label{subsec:apTrapezoidal}\renewcommand{\theequation}{A\arabic{equation}}\setcounter{equation}{0}
In this appendix, we prove a rigorous formula to write the integral in eq.~\eqref{eq:trape} as a sum of the trapezoidal term $S(\zeta)$, the correction term
$C(\zeta)$, and the error term $E(\zeta)$:
\begin{equation}\label{eq:appmoriI}
I \equiv \int_{-\infty}^{\infty} dt \, g(t) \equiv S(\zeta) + C(\zeta) + E(\zeta).
\end{equation}
where the error term $E(\zeta)$ is written as a contour integral over the integrand $\Psi(z) g(z)$:
\begin{equation} \label{eq:apppezeta}
E(\zeta) = \oint_C dz\, \Psi(z) g(z) ,
\end{equation}
and the trapezoidal sum $S(\zeta)$ is written as in eq.~\eqref{eq:trape}:
\begin{equation} \label{eq:apptrape}
S(\zeta) = \sum_{n=-\infty}^\infty \zeta g(n\zeta) . 
\end{equation}
Let's first take a look at the case $g(z)$ has no poles in the complex plane and $C(\zeta)=0$. Using Cauchy's residue theorem and choosing a path C such that all real
numbers in the real axis are included in its interior, the left hand side of eq.~\eqref{eq:appmoriI} can be written as:
\begin{eqnarray}
I &=& \int_{-\infty}^{\infty} dt\, \left[ \frac{1}{2\pi i} \oint_C dz\, g(z) \frac{1}{z-t}  \right] \nonumber \\
  &=& \frac{1}{2\pi i} \oint_C  dz\, g(z) \int_{-\infty}^{\infty} dt\, \frac{1}{z-t},  \label{eq:appIc} \end{eqnarray}
where we have assumed the validity of exchange of integration order. Similarly, we write the trapezoidal term as the following:
\begin{eqnarray}
S(\zeta) &=& \sum_{n=-\infty}^\infty \left[ \frac{1}{2\pi i} \oint_C dz\, g(z) \frac{\zeta}{z-n\zeta}  \right] \nonumber \\
         &=&  \frac{1}{2\pi i} \oint_C dz\, g(z) \sum_{n=-\infty}^\infty \frac{\zeta}{z-n\zeta}  \nonumber \\  
         &=& \frac{1}{2\pi i} \oint_C dz\, g(z) \pi \cot(\pi z/\zeta).
\label{eq:appSc} \end{eqnarray}
We have used a mathematical identity called Euler's partial fraction expression of the cotangent function:
\begin{eqnarray}
\pi \cot(\pi x) & \equiv &  \frac{1}{x} +\sum_{n=1}^\infty \left(  \frac{1}{x+n} + \frac{1}{x-n} \right) \nonumber \\
                &  =     &  \sum_{n=-\infty}^\infty \frac{1}{x-n} .
\label{eq:appeuler}\end{eqnarray}
Recalling another mathematical identity:
\begin{equation}\label{eq:appidinv}
\int_{-\infty}^\infty dt\, \frac{1}{z-t} = \left\{ \displaystyle  \begin{array}{cc} \displaystyle
                                                      -\pi i   &    {\rm Im}(z) > 0  \\  \displaystyle
                                                       \pi i   &    {\rm Im}(z) < 0   \end{array} \right. ,
\end{equation}
and subtracting $S(\zeta)$ from $I$, we have
\begin{eqnarray}
E(\zeta) & = &  I - S(\zeta)   \nonumber \\
     & = &  \frac{1}{2\pi i}  \oint_C dz\, g(z) \Psi(z) ,
\label{eq:appez} \end{eqnarray}
where
\begin{equation}\label{eq:apppsiz} 
 \Psi(z) = \frac{\displaystyle \mp 2\pi i}{\displaystyle 1-e^{\mp 2\pi i z/\zeta} } \mbox{\quad\quad} {\rm Im}(z)\gtrless 0 .
\end{equation}
which is eq.~\eqref{eq:char} in the main text. Direct evaluation of eq.~\eqref{eq:appez} along the two straight lines in Fig.~\ref{fig:contournopole} will yield an
error bound for the trapezoidal approximation to the integral. If there is a pole $z=i\omega$ on the positive imaginary axis for the integrand $g(z)$ as shown in
the bottom-left part of Fig.~\ref{fig:contourpoles}, we choose the contour line to be $C-C_+$ instead of $C$. Following steps from eq.~\eqref{eq:appIc}
to~\eqref{eq:appez}, we have
\begin{eqnarray}
 && E(\zeta) + C(\zeta)  =   I - S(\zeta) \nonumber \\
 &&   \quad\quad         =   \frac{1}{2\pi i} \oint_{C-C_+} dz\, g(z) \Psi(z)  \nonumber \\
 &&   \quad\quad         =   \frac{1}{2\pi i} \oint_C dz\, g(z)\Psi(z) - \frac{1}{2\pi i} \oint_{C_+} dz\, g(z) \Psi(z)  \nonumber \\
 &&   \quad\quad         =   E(\zeta) - {\rm Res}\left[ \Psi(z)g(z), i\omega \right]  ,
\label{eq:appez2}\end{eqnarray}
where ${\rm Res}\left[ \Psi(z)g(z), i\omega \right]$ denotes the residue of $\Psi(z)g(z)$ at the point $z= i\omega$. When
\begin{equation}\label{appgz2} 
g(z) = \frac{ e^{-(\omega^2+z^2)} e^{iz\nu} }{\omega^2+z^2} ,
\end{equation}
by choosing the contour as in the bottom-left part of Fig.~\ref{fig:contourpoles} we obtain the expression for the correction term:
\begin{equation}
C(\zeta) = -{\rm Res}\left[ \Psi(z)g(z), i\omega \right] = \frac{\pi}{\omega} \frac{e^{-\omega\nu}}{ 1- e^{2\pi\omega/\zeta} }, 
\end{equation}
which is eq.~\eqref{eq:residueu} in the main text.
%%%%%%%%%%%%%%   Appendix B   %%%%%%%%%%%%%%%%%%%%%%%%%%%%%%%%%%%%%%%%%%%%%%%%%%%%%%%%%%%%%%%%%%%%%%%%%%%%%%%%%%%%%%%%%%%%%%%%%%%%%%%%%%%%%%%%%%%%%%%%%%%%%%%%%%%%
\subsection{Error bounds for the trapezoidal approximation to the integral}
\label{subsec:errorbound}\renewcommand{\theequation}{B\arabic{equation}}\setcounter{equation}{0}
In this appendix, we discuss how to obtain eqs.~\eqref{eq:e0bu} and~\eqref{eq:e0bd} and eqs.~\eqref{eq:ehbu} and~\eqref{eq:ehbd}. Because the major part of $g^0(z)$
in eq.~\eqref{eq:g0z} is the limit of $\omega\to 0$ of $g^h(z)$ in eq.~\eqref{eq:ghz}, it is enough to consider the general case in eqs.~\eqref{eq:ehbu}
and~\eqref{eq:ehbd}. We first consider the error bound in eq.~\eqref{eq:ehbu}:
\begin{equation}\label{eq:appEu}
E^u = \left|  \frac{1}{2\pi i} \int_{z=\xi+i\theta} dz\, \Psi(z)g^h(z) \right|.
\end{equation}
Substituting $z=\xi+i\theta$, $\Psi(z)$ of eq.~\eqref{eq:char}, and $g^h(z)$ of eq.~\eqref{eq:ghz}, the left hand side of eq.~\eqref{eq:ehbu} can be evaluated as:
\begin{eqnarray}
E^u &=& \left| \int_{-\infty}^{\infty} d\xi\, \frac{e^{-(\omega^2 + (\xi+i\theta)^2)+i(\xi+i\theta)\nu  }} 
{ (1-e^{2\pi(\theta-i\xi)/\zeta})\left(\omega^2 + (\xi+i\theta)^2 \right) }  \right| \nonumber \\ 
&\leqslant&  \int_{-\infty}^{\infty} d\xi\, \frac{e^{-(\omega^2+\xi^2)} {e^{\theta^2-\theta\nu} } \left| e^{i(\xi\nu - 2\xi\theta)} \right| }
   {\left| 1 - e^{2\pi(\theta-i\xi)/\zeta} \right|\cdot \left|\omega^2+(\xi+i\theta)^2 \right|   }   \nonumber \\
&=& \int_{-\infty}^{\infty} d\xi\, \frac{e^{-(\omega^2+\xi^2)} {e^{\theta^2-\theta\nu} }}
   {\left| 1 - e^{2\pi(\theta-i\xi)/\zeta} \right|\cdot \left|\omega^2+(\xi+i\theta)^2 \right|   }   \nonumber \\
& < &  \int_{-\infty}^{\infty} d\xi\, \frac{e^{-(\omega^2+\xi^2)} {e^{\theta^2-\theta\nu} }}
   {\left| 1 - e^{2\pi(\theta-i\xi)/\zeta} \right|\cdot \left|\theta^2-\omega^2 \right|   }   \nonumber \\
& < &\int_{-\infty}^{\infty} d\xi\, \frac{e^{-(\omega^2+\xi^2)} {e^{\theta^2-\theta\nu} }}
   {\left( e^{2\pi\theta/\zeta} - 1 \right)\left| \theta^2-\omega^2 \right|   } \nonumber \\
& = & \frac{e^{-\omega^2} e^{(\theta-(\pi/\zeta+\nu/2))^2} e^{-(\pi/\zeta+\nu/2)^2}}  {\left| \theta^2-\omega^2 \right|(1-e^{-2\pi\theta/\zeta })} 
      \int_{-\infty}^{\infty} d\xi\, e^{-\xi^2}  \nonumber \\
& = & \frac{e^{-\omega^2}\sqrt\pi e^{(\theta-(\pi/\zeta+\nu/2))^2} e^{-(\pi/\zeta+\nu/2)^2}}  {\left| \theta^2-\omega^2 \right|(1-e^{-2\pi\theta/\zeta})}  ,
\label{eq:appehbu}
\end{eqnarray}
which is the right hand side of eq.~\eqref{eq:ehbu}. Following the detailed steps in eq.~\eqref{eq:appehbu} for the case of 
\begin{equation}\label{eq:appEd}
E^d = \left|  \frac{1}{2\pi i} \int_{z=\xi-i\beta} dz\, \Psi(z)g^h(z) \right|.
\end{equation}
we are able to easily prove eq.~\eqref{eq:ehbd}. By taking the limit of $\omega\to 0 $ in eqs.~\eqref{eq:ehbu} and~\eqref{eq:ehbd}, we then obtain
eqs.~\eqref{eq:e0bu} and~\eqref{eq:e0bd}.

%%%%%%%%%%%%%%%%%%%%%%%%%%%%%%%%%%%%%%%%%%%%%%%%%%%%%%%%%%%%%%%%%%%%%%%%%%%%%%%%%%%%%%%%%%%%%%%%%%%%%%%%%%%%%%%%%%%%%%%%%%%%%%%%%%%%%%%%%%%%%%%%%%%%%%%%%%%%%%%%
\section*{Acknowledgement}
This work was supported by NSFC grant no.91127015 and no.21103063  (Z. H.), the innovation project from the State Key Laboratory of Supramolecular Structure and
Materials (Z. H.), the open project from the State Key Laboratory of Theoretical and Computation Chemistry at Jilin University (Z. H.) and Graduate Innovation Fund
of Jilin University (project no.20121060). We gratefully acknowledge the Jilin University Supercomputing Center for providing resources for our work. We thank Ryan
Daly at the University of Iowa for reading our manuscript.
%%%%%%%%%%%%%%%%%%%%%%%%%%%%%%%%%%%%%%%%%%%%%%%%%%%%%%%%%%%%%%%%%%%%%%%%%%%%%%%%%%%%%%%%%%%%%%%%%%%%%%%%%%%%%%%%%%%%%%%%%%%%%%%%%%%%%%%%%%%%%%%%%%%%%%%%%%%%%%%%

%\bibliography{refs}

\providecommand{\latin}[1]{#1}
\providecommand*\mcitethebibliography{\thebibliography}
\csname @ifundefined\endcsname{endmcitethebibliography}
  {\let\endmcitethebibliography\endthebibliography}{}

%\newpage
%\begin{figure}[udtp]
%\begin{center}
%\centerline{
%\psfig{file=toc.ps,width=3.5in,angle=270}}
%\caption{TOC graphics}
%\end{center}
%\label{fig:toc}
%\end{figure}

\end{document}